\documentclass[pre,aps,twocolumn,showpacs,preprintnumbers,superscriptaddress]{revtex4}

\usepackage{graphicx}
\usepackage{dcolumn}
\usepackage{bm}
\usepackage{amssymb}
\usepackage{pifont}
\usepackage{amsmath}
\usepackage{times}
\usepackage[nooneline]{subfigure}

\begin{document}

\title{Dynamics of perturbations in disordered chaotic systems}


\author{Ivan G. Szendro} \email{szendro@ifca.unican.es}
\altaffiliation[Present address: ]{Max Planck Institute for the
Physics of Complex Systems, N\"{o}thnitzer Strasse 38, 01187 Dresden,
Germany}
\affiliation{Instituto de F\'{\i}sica de Cantabria (IFCA),
CSIC--UC, E-39005 Santander, Spain}

\affiliation{Departamento de F{\'\i}sica Moderna, Universidad de
Cantabria, Avda. Los Castros, E-39005 Santander, Spain}

\author{Juan M. L{\'o}pez} \email{lopez@ifca.unican.es}
\affiliation{Instituto de F\'{\i}sica de Cantabria (IFCA),
CSIC--UC, E-39005 Santander, Spain}

\author{Miguel A. Rodr{\'\i}guez} \email{rodrigma@ifca.unican.es}
\affiliation{Instituto de F\'{\i}sica de Cantabria (IFCA),
CSIC--UC, E-39005 Santander, Spain}

\begin{abstract} 
We study the time evolution of perturbations in spatially extended chaotic systems in the
presence of quenched disorder. We find that initially random perturbations tend to
exponentially localize in space around static pinning centers that are selected by the
particular configuration of disorder. The spatial structure of typical perturbations,
$\delta u(x,t)$, is analyzed in terms of the Hopf-Cole transform, $h(x,t) \equiv \ln
\vert\delta u(x,t)\vert$. Our analysis shows that the associated surface $h(x,t)$
self-organizes into a faceted structure with scale-invariant correlations. Scaling
analysis of critical roughening exponents reveals that there are three different
universality classes for error propagation in disordered chaotic systems that correspond
to different symmetries of the underlying disorder. Our conclusions are based on numerical
simulations of disordered lattices of coupled chaotic elements and equations for diffusion
in random potentials. We propose a phenomenological stochastic field theory that gives
some insights on the path for a generalization of these results for a broad class of
disordered extended systems exhibiting space-time chaos.
\end{abstract}

\pacs{05.45.Jn, 05.45.Ra, 05.40.-a}

\maketitle 

\section{Introduction} Spatially extended chaotic systems (SECS) are of great importance
for the understanding of fundamental problems in deterministic many-particle systems,
including hydrodynamics and turbulence~\cite{bohr,frisch} or weather
forecasting~\cite{kalnay}. These systems exhibit dynamical instabilities that can be
quantified by the spectrum of Lyapunov exponents and their corresponding Lyapunov
vectors (LVs)~\cite{bohr,eckmann,ott,legras96}.

Recently, the application of some tools and concepts borrowed from the physics of
nonequilibirum statistical systems has been shown to be a very promising line of research
to study certain aspects of SECS. In particular, it has been observed that (after a
suitable logarithmic transformation) the evolution of an infinitesimal perturbation in
deterministic SECS can be generically described as a scale-invariant rough
surface~\cite{pikovsky_roughening_1994,pikovsky_dynamic_1998,pikovsky_dynamic_2001,
sanchez_rare_2004,szendro_CLV07}. In this surface picture, the erratic fluctuations of the
system, due to the chaotic dynamics, are interpreted as noise. Remarkably, it has been
shown~\cite{pikovsky_roughening_1994,pikovsky_dynamic_1998,szendro_CLV07} that in many
cases the associated surfaces belong to the Kardar-Parisi-Zhang (KPZ)~\cite{kardar1986}
universality class of nonequilibrium surface roughening. Nonetheless, the probability
distribution of the randomness generated by the chaotic trajectory can be crucial. In this
regard, the existence of long-range correlations~\cite{pikovsky_dynamic_2001} or a fat
tail of the distribution~\cite{sanchez_rare_2004} may change the universality class
observed in the surface growth picture.

Typically the above discussed studies deal with {\em homogeneous} systems, where only
identical elements are coupled, often in a diffusive manner. However, much less is known
about the evolution of chaotic perturbations in {\em inhomogeneous} extended systems,
where the many coupled elements are either different or the coupling itself is a quenched
random field that varies along the system. Such conditions naturally arise in
some important applications, {\it e.~g.}~in regional weather
forecasting, where inhomogeneity is present in the form of explicit surface-topography
dependent terms~\cite{kalnay,charney,pedlosky,legras_ghil} in the dynamic equations.

Inhomogeneous SECS can indeed demonstrate rather unusual properties, like for instance
taming of spatiotemporal chaos induced by disorder~\cite{brainman}, disorder-enhanced
synchronization~\cite{wiesenfield,mousseau}, and avoided crossing and level repulsion
similar to that ocurring for energy eigenvalues in disordered quantum
systems~\cite{ahlers_zillmer_pikovsky}. 

In this paper, we focus on the dynamics of perturbations in disordered SECS, which has
been little investigated in the literature. We analyze here a rather simple but
enlightening model consisting of a lattice of coupled chaotic elements whose parameters
are randomly distributed or the coupling among them is a quenched random variable. We show
that long-range temporal correlations, induced by the quenched disorder, can cause
perturbations to strongly localize. The strong localization observed here is essentially
different from the dynamical localization that has been previously reported for
homogeneous SECS~\cite{pikovsky_roughening_1994,szendro_CLV07}. In the latter, the
positions of the localization centers keep fluctuating in space, while in the presence of
quenched disorder we find that the positions of the localization centers are fixed for a
given disorder realization. We analyze typical perturbations, $\delta u(x,t)$, by making
use of the Hopf-Cole transform, $h(x,t) \equiv \ln\vert\delta u(x,t)\vert$, and mapping to
the equivalent surface growth problem. Strong localization leads to the formation of
faceted structures in the corresponding surface picture. These Lyapunov surfaces also
exhibit a coarsening behavior and anomalous kinetic roughening, akin to some
nonequilibrium growing surfaces. The connection of the problem of the propagation of
errors in disordered SECS with the problem of diffusion in random potentials is also
discussed. We propose a phenomenological stochastic field theory pointing toward a
generalization of our results for a broad class of disordered extended systems exhibiting
space-time chaos.

\section{Description of the models} We focus our numerical study on coupled-map lattices
with quenched disorder. We consider $L$ coupled chaotic maps $u_x(t)$, with $x = 1, 2,
\cdots L$, following the evolution equation
\begin{eqnarray}
u_x(t+1)&=& \sigma_{x,x+1}
f_{x+1}(u_{x+1}(t))+\sigma_{x,x-1} f_{x-1}(u_{x-1}(t))\nonumber\\
&&+(1 - 2\sigma_{x,x} )f_x(u_x(t)) ,\nonumber
\end{eqnarray}
where the $\sigma_{x,x \pm 1}$ are the disordered nearest-neighbor coupling constants,
$\sigma_{x,x}$ is the on-site contribution, and $f_x$ is the local nonlinear map at site
$x$. Here we choose $f_x(\varrho)=c_x(1/2-|\varrho\mod 1-1/2|)$, which is a periodic
continuation of the tent-map. This map shows chaotic behavior for $c_x\in (1,\infty)$,
with a Lyapunov exponent that grows logarithmically with $c_x$. This choice allows one to
study extended coupled systems with a spatial distribution of Lyapunov exponents by
allowing $c_x$ to take randomly distributed values along the lattice.

Almost any initial random perturbation $\delta u_x(t=0)$ will grow in magnitude and
develop space-time correlations while propagating along the system, quickly aligning with
the most unstable direction in tangent space, the so-called main LV:
\begin{eqnarray}
\delta u_x(t+1) &=& \sigma_{x,x+1}
f^\prime_{x+1}(u_{x+1}(t))\delta
u_{x+1}(t)\nonumber\\
&&+\sigma_{x,x-1}
f^\prime_{x-1}(u_{x-1}(t))\delta u_{x-1}(t) \label{evoerrors}\\
&&+(1 - 2\sigma_{x,x} )f^\prime_x(u_x(t))\delta
u_x(t),\nonumber
\end{eqnarray}
where $f_x^\prime(\varrho)$ is just the derivative of the local map $f_x(\varrho)$
with respect to its argument $\varrho$. 

Numerical integration of the tangent space equations (\ref{evoerrors}) looks apparently
rather simple, however, a caveat is in order. We found that the introduction of quenched
disorder leads to the appearance of very large differences among error field $\delta
u_x(t)$ values at certain sites of the system. The reason for this is the quenched nature
of the disorder so that, if a large value of the random variable is assigned to a given
site, it will continue giving high contributions for all times. In fact, these differences
can become so large during the simulation that, if we were to naively integrate
(\ref{evoerrors}) and just multiply by some global factor to avoid numerical overflow, the
perturbation field at those sites where the perturbation values are small would soon be
considered as zero by the computer due to accuracy limitations. It is rather simple to
overcome this technical problem by avoiding calculating the $\delta u_x(t)$ directly, but
rather computing the quotients $\vartheta_x(t+1)=\delta u_x(t+1)/\delta u_x(t)$ and
$\varphi_x(t)=\delta u_{x+1}(t)/\delta u_x(t)$ instead. We therefore rewrite
(\ref{evoerrors}) as
\begin{eqnarray}
\vartheta_x(t+1)&=&\sigma_{x,x+1}
f^\prime_{x+1}(u_{x+1}(t))\varphi_{x}(t)\nonumber\\
&&+\sigma_{x,x-1} f^\prime_{x-1}(u_{x-1}(t))/\varphi_{x-1}(t)\nonumber\\
&&+(1 - 2\sigma_{x,x})f^\prime_x(u_x(t))\nonumber\\
\varphi_x(t+1)&=&\varphi_x(t)\frac{\vartheta_{x+1}(t+1)}{\vartheta_x(t+1)}.\nonumber
\end{eqnarray}
Now we can numerically integrate this pair of equations instead of Eq.~(\ref{evoerrors}).
This useful numerical trick resolves the problem of possible overflows during the
simulations.

We have studied three different scenarios for a disordered couple-map lattice system. In
the first scenario (model A) we consider the case in which the couplings are identical
along the system, but coupled elements are inhomogeneous. In the simplest setting we can
model this situation by drawing the local map constant $c_x$ from a uniform distribution,
so we have
\begin{equation}			
\begin{array}{ll} \text{Model
A:}&\left\{
\begin{array}{lll}c_x&\in&\mathrm{U}(1,2)\\\sigma_{x,y}&=&1/3
\end{array}, \right.
\end{array} \label{model1cq}
\end{equation}
where $y\in\{x-1,x,x+1\}$. Note that we chose a homogeneous democratic coupling
($\sigma_{x,y}= 1/3$) all along the system. In this situation individual elements are more
chaotic or less chaotic depending on the corresponding value of $c_x$. Larger values of
$c_x$ lead to a locally faster growth of the perturbation at site $x$. Strictly speaking,
this is only true for uncoupled maps. Things are a much more involved here since the maps
are coupled. Thus, the behavior of the perturbation at one site does not only depend on
the map at that specific site, but also on the dynamics of its neighborhood.

\begin{figure}
\centerline{\includegraphics *[width=0.40\textwidth]{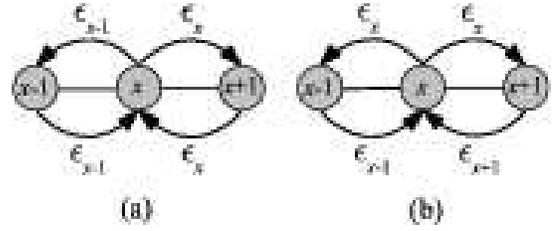}}
\caption{
The coupling configurations of one site to its neighboring sites
and viceversa for (a) the random barrier (RB)
and (b) random trap (RT) model.\label{fig1}}
\end{figure}

A different source of disorder we have explored is the existence of a quenched random
coupling, while the individual maps are all identical. This corresponds to systems with a
disordered diffusion coefficient. From the classical theory of diffusion in disordered
media we know that two different symmetries are of
interest~\cite{haus_diffusion_1987,bouchaud_anomalous_1990}. These two models are sketched
in Fig.~\ref{fig1} and described in detail below.

On the one hand, we have the {\em random barrier} (RB)
model~\cite{haus_diffusion_1987,bouchaud_anomalous_1990}. Physically this corresponds to
the existence of a random potential barrier between every two neighboring sites so that
the diffusion coefficient from site $x$ to site $x+1$ is equal to that from site $x+1$ to
site $x$ (see Fig.~\ref{fig1}). In other words, the disorder is associated with the bonds
connecting neighboring sites on the lattice. We call this configuration model B: 
\begin{equation}
\begin{array}{ll} \text{Model B:} &\left\{
\begin{array}{lll}c_x&=&2\\ 
\sigma_{x,x-1}&=&\epsilon_x\\
\sigma_{x,x+1}&=&\epsilon_{x+1}\\
\sigma_{x,x}&=&(\epsilon_{x}+\epsilon_{x+1})/2\end{array},\right.
\end{array} \label{model2cq}
\end{equation}
where the disorder is drawn from a uniform distribution,
$\epsilon_{x}\in\mathrm{U}(0,\epsilon_0)$, and $\epsilon_0$ being an arbitrary
parameter
that we take as $1/3$ unless otherwise stated.

On the other hand, some disordered systems are better described by the so-called {\em
random trap} (RT) model~\cite{haus_diffusion_1987,bouchaud_anomalous_1990}. In this case a
static disordered diffusion coefficient is assigned to each lattice site $x$ instead to
bonds. This corresponds to our model C, which is defined by 
\begin{equation}
\begin{array}{ll} \text{Model C:} &\left\{
\begin{array}{lll}c_x&=&2\\ 
\sigma_{x,x-1}&=&\epsilon_{x-1}\\
\sigma_{x,x+1}&=&\epsilon_{x+1}\\
\sigma_{x,x}&=&\epsilon_{x}\end{array},\right.
\end{array} \label{model3cq}
\end{equation}
where the disorder is uniformly distributed,
$\epsilon_{x}\in\mathrm{U}(0,\epsilon_0)$,
and again $\epsilon_0$ can be any number in the interval $(0,1/2)$, here
we use $\epsilon_0=1/3$.

\begin{figure}
\centerline{\includegraphics *[width=0.40\textwidth]{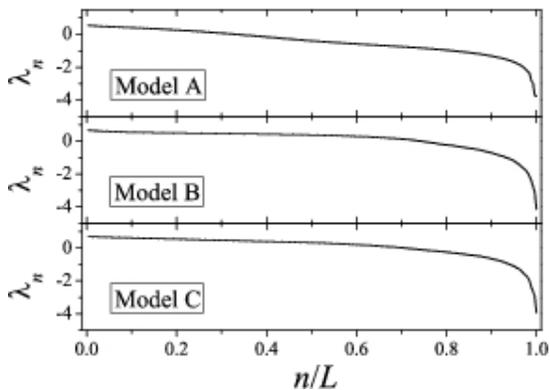}}
\caption{
Lyapunov spectra for models A, B and C in a lattice of $L=256$
coupled tent maps after average over ten different disorder realizations.\label{fig2}}
\end{figure}

Typical Lyapunov spectra corresponding to the three disordered models introduced here are
shown in Fig.~\ref{fig2} for a system of size $L=256$ and averages over 10 different
disorder realizations. One can see that, depending on the model, one has 30-80~\% of the
spectra in the region $\lambda > 0$. Note that Lyapunov spectra are self-averaging,
therefore spectra for a single realization differ from the average in small fluctuations,
which for large enough systems are negligible.

Some remarks are now in order. First, let us stress that these two disordered
diffusion configurations, namely RB and RT, were not chosen arbitrarily, but are in fact
all existing physically meaningful ways to introduce disorder in the
diffusive couplings~\cite{haus_diffusion_1987,bouchaud_anomalous_1990}. Second, it is
important to remark that in this paper we focus on systems in the presence of
{\em weak} disorder, where the probability density of having zero disorder at any given
site, ${\cal P}(\epsilon=0)$, is bounded. We do leave out of our study the case of {\em
strong} disorder, ${\cal P}(\epsilon \to 0) \sim \epsilon^{-\vert\nu\vert}$,
where a fraction of the system sites can effectively act as sinks, which is known to
have a great impact on the asymptotic transport
properties~\cite{haus_diffusion_1987,bouchaud_anomalous_1990} and will be discussed
elsewhere. 

In principle, classical results on the problem of diffusion in random
media~\cite{haus_diffusion_1987,bouchaud_anomalous_1990} may also be invoked to argue that
the details of the disorder distribution should be irrelevant in the sense that weak
disorder gives rise to the same asymptotic dynamical behavior independently of the
detailed form of the disorder distribution. The same conclusion applies in the case of
strong disorder, where the asymptotic dynamics is fully determined by the form of the
divergence at the origin, {\it i.e.}~the value of the exponent $\nu$. In the following we
focus on the case of weak disorder and a uniform distribution ${\cal P} (\epsilon) =
\mathrm{U}(0,\epsilon_0)$ with the parameter $\epsilon_0=1/3$ is used to exemplify our
results. However, in the presence of random multiplicative terms the critical behavior
might be affected by the detailed form of the disorder distribution~\cite{krug1993}.

\section{Surface evolution and coarsening} As occurs in the case of homogeneous systems,
the dynamics of the main LV can be conveniently described in terms of the equivalent
surface
picture~\cite{pikovsky_roughening_1994,pikovsky_dynamic_1998,pikovsky_dynamic_2001,
sanchez_rare_2004,szendro_CLV07}. The reason is that the logarithm of the perturbation
field turns out to be scale-invariant, so that correlations have the form of power-law
functions and critical exponents can then be used to characterize the space-time structure
of perturbations. Therefore, in this paper we are interested in the scaling properties of
the rough surface defined by the Hopf-Cole transform of the perturbation $h(x,t) \equiv
\ln{|\delta u_x(t)|}=\ln{|\delta u_x(0)|}+\sum_{\tau=1}^t\ln{|\vartheta_x(\tau)|}$.

\begin{figure}
\centerline{\includegraphics *[width=0.40\textwidth]{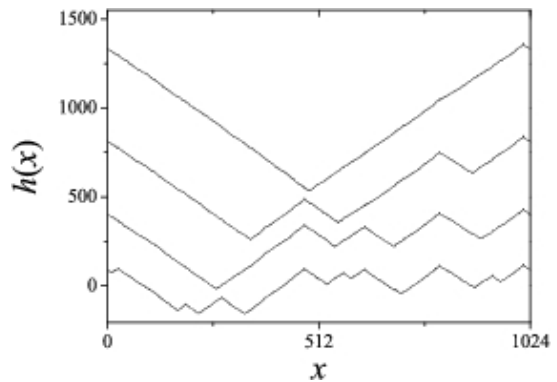}}
\caption{Typical evolution a Lypunov vector surface $h(x,t)$.
One can observe the emergence of facets that grow in time until, in
the long time limit, the whole interface is spanned by one single
facet. Tiny fluctuations can be observed at small scales. The data
correspond to model C in a lattice of $L=1024$ maps.\label{fig3}}
\end{figure}

Space-time scaling properties of chaotic perturbations in inhomogeneous systems turn out
to be very different from those in homogeneous systems, even in the case of weak disorder
studied here. As can be immediately seen in Fig.~\ref{fig3}, a first observation is the
patterned structure of the LV surface, which is visible to the naked eye and contrasts
with the KPZ-type morphologies observed in homogeneous
systems~\cite{pikovsky_roughening_1994,szendro_CLV07}. This already indicates that
significant differences are to be expected in the scaling properties of inhomogeneous
systems. Even weak disorder induces self-organization in a triangular structure, which
reflects the strong spatial localization of the perturbation in the form of a exponential
profile around some strong pinning centers. Note that a facet, $h(x,t) = h(x_0) - s(t)
|x-x_0|$, with a cusp at $x=x_0$ and slope $s(t)$ corresponds to an exponential profile of
the perturbation, $\delta u_x(t) \propto \exp [-s(t)|x-x_0|]$, around the pinning center
$x_0$.

In Fig.~\ref{fig3} we plot a typical evolution of the main LV according to
Eq.~(\ref{evoerrors}) in a system of size $L=1024$ for a given disorder realization in the
case of model C. We observe that at short times, the surface $h(x,t)$ is composed of
triangular facets of varying sizes which grow in time until, in the long time limit, the
whole interface is formed by just one single facet. Therefore, as time evolves the size of
the triangles increases and number of triangles diminishes. This surface dynamics
corresponds to non-equilibrium 'coarsening'. Similar triangular morphologies are obtained
for the three models A, B and C introduced above. Close inspection and analysis of
critical exponents shows that the three models actually belong to different universality
classes.

\begin{figure}
\centerline{\includegraphics *[width=0.40\textwidth]{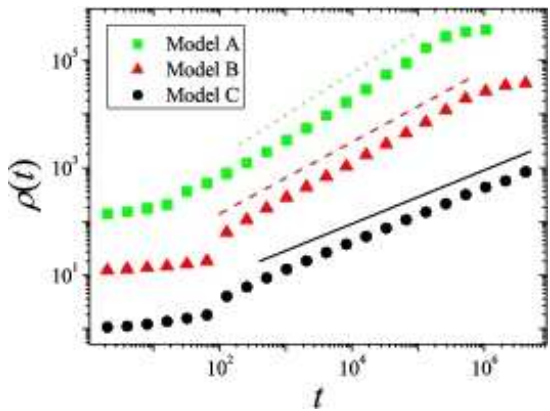}}
\caption{(Color online) The evolution of the lateral mound size
$\rho(t)$ is plotted for the three models. The straight lines are plotted to
guide the eye and have slopes $0.78$ (dotted), $2/3$ (dashed) and
$1/2$ (solid), respectively. The curves for models A and B
have been shifted for better visibility. \label{fig4}}
\end{figure}

In order to characterize the growth of the faceted pattern we have measured the lateral
mound size $\rho(t)$, which gives the coarsening length or typical length scale of the
instability. This is usually done by calculating the slope-slope correlation function
$\langle \overline{\nabla h(x,t) \nabla h(x+r,t)} \rangle$ and measuring the distance,
$\rho(t)$, at which this correlation crosses zero. We calculate the surface gradient as
the centered discrete derivative $\nabla h(x,t)\equiv [h(x+1,t)-h(x-1,t)]/2$. In
Fig.~\ref{fig4} the lateral mound size, $\rho(t)$, for the three models is shown. After a
short initial period, $\rho(t)$ grows as a power law in time with exponents
$\theta_\mathrm{B}=0.67\pm 0.03$ and $\theta_\mathrm{C}=0.48\pm 0.03$, for models B and C,
respectively. At long times $\rho(t)$ saturates, as the surface becomes dominated by a
single facet and the coarsening length becomes comparable to the system size. For model A
the intermediate region does not seems to be following a simple power-law. Nonetheless, we
have fitted an exponent $\theta_\mathrm{A}=0.78\pm 0.05$ for the sake of comparison. The
different scaling exponents for the coarsening length already indicate that the three
models possibly belong to different universality classes.

\section{Surface roughness scaling} Faceted surfaces exhibit anomalous scaling
properties in the sense that local and global fluctuations may scale with different
scaling functions (and exponents), which can be cast in the generic dynamic scaling form
introduced by Ramasco {\it et.~al.}~\cite{ramasco_generic_2000}. In this section we
briefly describe the generic scaling theory for surface kinetic roughening that will be
used later on to analyze our numerical data.

The roughness of scale-invariant surfaces corresponds to the fluctuations of
the surface height. These fluctuations can be computed either locally or globally as
follows. On the one hand, the global roughness exponent $\alpha$ can be obtained from the
scaling behavior of the global width
$W(L,t)=\langle\overline{[h(x,t)-\overline{h}(t)]^2}\rangle^{1/2}$. Here, the overline
denotes an average over all sites $x$ in a system of size $L$ and brackets denote the
average over different realizations. For scale-invariant surfaces one expects the global
width to scale as $W(L,t)= t^{\alpha/z} \mathcal{G}(L/t^{1/z})$, where $\mathcal{G}(u)$ is
a scaling function that becomes constant for $u\gg 1$ and decays as $\sim u^{\alpha}$ for
$u\ll 1$. The roughness exponent $\alpha$ and dynamic exponent $z$ characterize the
scaling behavior of the global surface fluctuations.

On the other hand, one can measure the local roughness exponent $\alpha_\mathrm{loc}$,
which is defined via the scaling behavior of the local width
$w(l,t)=\langle\langle[h(x,t)-\langle h\rangle_l(t)]^2\rangle_l\rangle^{1/2}$, where
$\langle \ldots\rangle_l$ denotes an average over $x$ in a window of size $l$. The local
width scales as $w(l,t)\sim t^{\alpha/z}\mathcal{G}_\mathrm{A}(l/t^{1/z})$, where the
scaling function $\mathcal{G}_\mathrm{A}(u)$ has a similar asymptotic behavior as
$\mathcal{G}(u)$, but with a local anomalous exponent, $\sim u^{\alpha_\mathrm{loc}}$ for
$u
\ll 1$. So that in the stationary regime (for $t \gg L^z$) one obtains
$w_\mathrm{stat}(l,L)\sim l^{\alpha_\mathrm{loc}}L^{\alpha-\alpha_\mathrm{loc}}$. In the
cases where local and global roughness exponents do not coincide the scaling is said to be
{\em anomalous}.
\begin{figure}
\centerline{\includegraphics *[width=0.40\textwidth]{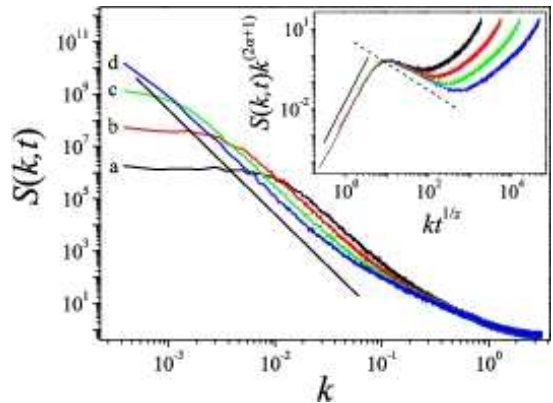}}
\caption{(Color online) The spectral power functions of the surfaces obtained for
model A are plotted for times
$t=2^{12}(\mathrm{a}),2^{14}(\mathrm{b}),2^{16}(\mathrm{c}),2^{18}(\mathrm{d})$.
The straight line corresponds to $\alpha_s=1.5$. In the inset we
show the data collapse according to the scaling ansatz
(\ref{ramascoscale}) and (\ref{ramscoscalef}), using $\alpha=1.03$
and $z=1.3$. The dashed and the solid straight line have slopes -1
and 3, respectively. The plotted data correspond to model A for a
system of size $L=16384$ averaged over 100 disorder realizations.\label{fig5}}
\end{figure}

Following Ramasco {\it et.~al.}~\cite{ramasco_generic_2000}, in order to correctly
classify the different forms that scaling can take it is convenient to introduce a third
roughness exponent, namely the so called spectral roughness exponent $\alpha_\mathrm{s}$.
This is defined in terms of the structure factor (or power spectrum in $k$ space),
\begin{equation}
S(k,t)=\langle \widehat{h}(k,t) \widehat{h}(-k,t)\rangle,
\end{equation}
where $\widehat{h}(k,t)$ is the spatial Fourier transform of $h(x,t)- \overline{h}(t)$.
Ramasco {\it et.~al.}~\cite{ramasco_generic_2000} showed that scale-invariant roughening
in $d+1$ dimensions is fully described in the following scaling ansatz:
\begin{equation}
S(k,t)=k^{-(2\alpha + d)}s(kt^{1/z}),\label{ramascoscale}
\end{equation}
$z$ being the dynamical exponent that connects temporal-scales $\tau$ and length-scales
$l$ according to $\tau \sim l^z$, and the nontrivial scaling function takes the form
\begin{equation}
s(u)\sim \left\{ \begin{array}{lll}
u^{2(\alpha-\alpha_\mathrm{s})} & \text{if}&  u \gg 1
\\ u^{2 \alpha + d} & \text{if}& u \ll 1.
\end{array}\right.\label{ramscoscalef}
\end{equation}
Here $\alpha$ is the global roughness exponent. With the help of these three exponents one
can distinguish the four different possible types of scaling that any scale-invariant
rough surface can exhibit,
\begin{equation}
\left\{ \begin{array}{llll}
\mathrm{if}&\alpha_\mathrm{s}<1\Rightarrow\alpha_\mathrm{loc}=\alpha_\mathrm{s}&\bigg\{
\begin{array}{ll}\alpha_\mathrm{s}=\alpha\Rightarrow&\text{Family-Vicsek}\\
\alpha_\mathrm{s}\neq\alpha\Rightarrow& \text{intrinsic}
\end{array} \\
\mathrm{if}&\alpha_\mathrm{s}>1\Rightarrow\alpha_\mathrm{loc}=1&\bigg\{
\begin{array}{ll}\alpha_\mathrm{s}=\alpha\Rightarrow&\text{super-rough}\\\alpha_\mathrm{s}
\neq\alpha\Rightarrow&
\text{faceted.}\end{array}
\end{array}\right. \label{sctype}
\end{equation}
Note that only in the case where standard Family-Vicsek scaling is valid, which means
$\alpha = \alpha_\mathrm{loc}= \alpha_\mathrm{s}$, the surface is self-affine. In the
other cases this will in general not be true, {\it i.~e.}~local and global fluctuations
will still exhibit dynamical scaling, but will do so with different exponents.

Although Family-Vicsek scaling is the best known scaling class
\cite{barabasi_fractal_1995}, over the last ten years experimental studies in a variety of
systems including growth of thin-films, electrodeposition, fracture or fluid imbibition
\cite{yang_instability_1994,jeffries_instability_1996,huo_anomalous_2001,
santamaria_scaling_2002,lopez_anomalous_1998,morel_anomalous_1998,soriano_anomalous_2002}
as well as theoretical studies
\cite{ramasco_generic_2000,amar_groove_1993,schroeder_scaling_1993,das_sarma_kinetic_1994,
das_sarma_scale_1996,smilauer_crossover_1994,bhattacharjee_infrared_1996,
dasgupta_controlled_1996,lopez_power_1997,lopez_superroughening_1997,krug_turbulent_1994,
lopez_scaling_1999,lopez_scaling_2005} have confirmed the existence of intrinsic anomalous
and super-rough kinetic roughening, as well as the validity of the generic scaling
theory.
\begin{figure}
\centerline{\includegraphics *[width=0.40\textwidth]{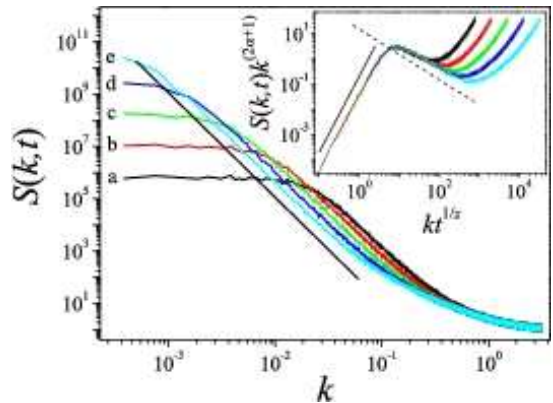}}
\caption{(Color online) The spectral power functions of interfaces
obtained for model B are plotted for times
$t=2^{12}(\mathrm{a}),2^{14}(\mathrm{b}),2^{16}(\mathrm{c}),2^{18}(\mathrm{d}),2^{20}
(\mathrm{e})$.
The straight line corresponds to $\alpha_s=1.5$. In the inset we
show the data collapse according to scaling ansatz
(\ref{ramascoscale}) and (\ref{ramscoscalef}), using $\alpha=1.0$
and $z=1.5$. The dashed and the solid straight line have slopes -1
and 3, respectively. The plotted data corresponds to model B for a
system of size $L=16384$ averaged over 100
realizations.\label{fig6}}
\end{figure}

\begin{figure}
\centerline{\includegraphics *[width=0.40\textwidth]{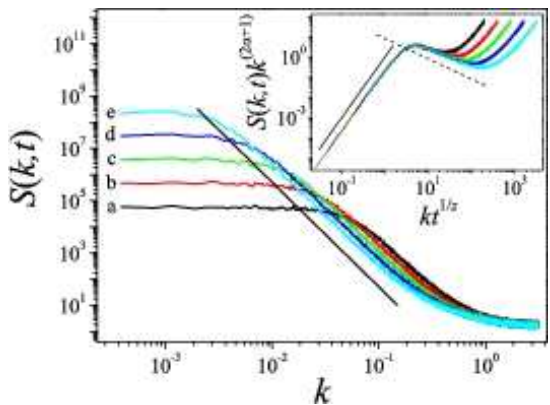}}
\caption{(Color online) The spectral power functions of interfaces
obtained for model C are plotted for times
$t=2^{12}(\mathrm{a}),2^{14}(\mathrm{b}),2^{16}(\mathrm{c}),2^{18}(\mathrm{d}),2^{20}
(\mathrm{e})$.
The straight line corresponds to $\alpha_s=1.5$. In the inset we
show the data collapse according to scaling ansatz
(\ref{ramascoscale}) and (\ref{ramscoscalef}), using $\alpha=1.0$
and $z=2.0$. The dashed and the solid straight line have slopes -1
and 3, respectively. The plotted data corresponds to model C for a
system of size $L=16384$ averaged over 100
realizations.\label{fig7}}
\end{figure}

We now use this general scaling ansatz to study the scaling properties of our system. In
order to do so, we have calculated the structure factor corresponding to the Hopf-Cole
transforms of the perturbations at various times. Figures~\ref{fig5}, \ref{fig6} and
\ref{fig7} show the structure factor for models A, B and C, respectively. We observe a
power-law decay as $\sim k^{-(2\alpha_\mathrm{s}+1)}$ for momenta limited by a large scale
and a small scale cutoff. At large length scales the power-law behavior is limited by the
correlation length $\xi(t)\sim t^{1/z}$, as expected from dynamical scaling. The cutoff at
short lengths also grows in time, suggesting the existence of a different dynamics at
small scales, which will be studied in detail in Sec.~V. Note that the curves $S(k,t)$
shift downward for increasing times, indicating the presence of anomalous scaling with
$\alpha_\mathrm{s}>\alpha$. Specifically, one expects $S(k,t) \sim
k^{2\alpha_\mathrm{s}+1}t^{(\alpha-\alpha_\mathrm{loc})/z}$. We have measured
$\alpha_\mathrm{s} = 1.43\pm 0.05$, $1.47\pm 0.05$ and $1.45\pm 0.05$ for models A, B and
C, respectively. This should be compared with the value of $\alpha_{\mathrm{s}}=3/2$ that
can be obtained analytically for interfaces consisting of random smooth facets
\cite{ramasco_generic_2000}. In the insets of Figs.~\ref{fig5}, \ref{fig6} and \ref{fig7}
we show a data collapse according to the generic scaling ansatz in
Eqs.~(\ref{ramascoscale}) and (\ref{ramscoscalef}), obtaining values for the roughness and
dynamical exponents $\alpha$ and $z$. As can be verified, all three models show scaling
properties well described by the faceted interface scaling class in Ramasco's
classification, Eq.~(\ref{sctype}), with a roughness exponent $\alpha\approx 1$ for all
the three models. Nonetheless, all three systems yield different values for the dynamical
exponent, $z_\mathrm{A}=1.3\pm 0.1$, $z_\mathrm{B}=1.50\pm 0.03$ and $z_\mathrm{C}=2.00\pm
0.03$, again indicating that the three models belong to different universality classes.
These values of $z$ define the typical length scale that correlations have spread along
the system up to time $t$, $\xi(t) \sim t^{1/z}$, and are in good agreement with the
inverse of the exponents ($\theta = 1/z$) that we obtained from the analysis of the
coarsening length discussed in Sec.~III for all the three models studied in this paper.

\section{Separating facets from fluctuations} The scaling analysis of the surface
fluctuations presented in the preceding sections has shown that the LV surface
self-organizes in a characteristic triangular pattern for all the three models of
disorder. A closer inspection, in particular the computation of the structure factor, has
also revealed the existence of a cutoff at small scales such that length scales below that
point obey a different dynamics. This suggests the existence of another dynamical process
taking place at small scales which is different from the mechanism responsible for the
large scale faceted structure.
\begin{figure}
\centerline{\includegraphics *[width=0.40\textwidth]{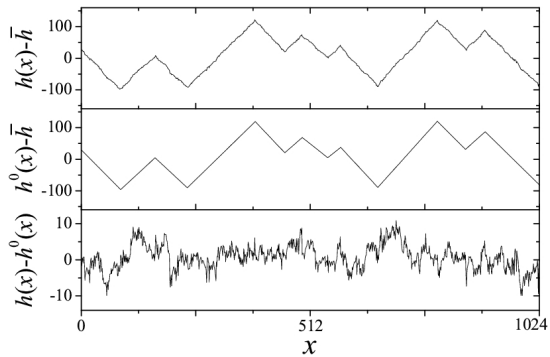}}
\caption{For a segment of 1024 maps on a system of $L=16364$ we
exemplify the separation of the surface $h(x)$ into the
triangular pattern $h^0(x)$ and the difference
$y(x) = h(x)-h^0(x)$, which gives the small scale fluctuations.\label{fig8}}
\end{figure}

In this section we show that one can actually separate two different contributions to the
LV surface height: The large scale pattern, $h^0(x,t)$, formed by facets of constant
slope, and a randomly fluctuating term, $y(x,t)$, which becomes the dominant one at
small scales. Therefore the surface profile at any given time can be expressed as the sum
of the two independent components as $h(x,t) = h^0(x,t) + y(x,t)$. This separation can be
carried out numerically as follows. For a given time the interface $h(x,t)$ is smoothed to
remove local maxima and minima corresponding to local fluctuations upon the triangular
structure. This is done by replacing the heights at every site by a spatial average over
some arbitrary region around each site. Then, the locations of the cusp sites and valley
sites of the smoothed curve are easily identified and the faceted pattern $h^0(x)$ is
defined as the set of straight lines connecting the cusp points to the neighboring valley
points. The interface corresponding to the fluctuations is then obtained by taking the
difference $y(x,t)=h(x,t)-h^0(x,t)$ as exemplified in Fig.~\ref{fig8}

\begin{figure}
\centerline{\includegraphics *[width=0.40\textwidth]{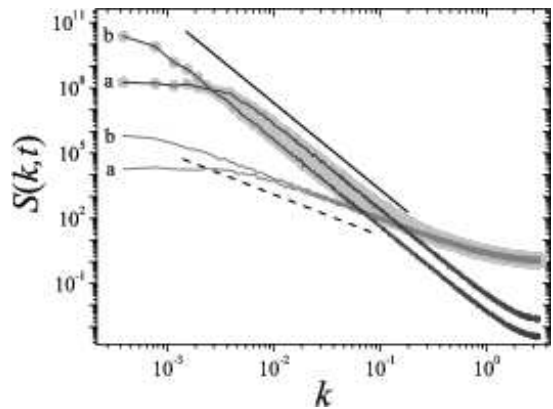}}
\caption{The power spectra of the interface $h(x)$ (light gray
dots), the triangular structure $h^0(x)$ (dark gray lines), and the
difference $y(x) = h(x)-h^0(x)$ (gray lines) are plotted for two
different times $t_\mathrm{a}<t_\mathrm{b}$. The straight solid
line corresponds to $\alpha^0_\mathrm{s}=1.5$, while the straight
dashed line corresponds to $\alpha^\mathrm{diff}_\mathrm{s}=0.45$.
\label{fig9}}
\end{figure}

Interestingly, one can observe that the triangular pattern, $h^0$, and the random
fluctuation components, $y(x,t)$, are uncorrelated. This can be shown by checking that
$\langle\widehat{h^0}(k,t) \widehat{y}(-k,t)\rangle=0$, which is equivalent to proving the
identity
\begin{equation}
S(k,t) = \langle\widehat{h^0}(k,t) \widehat{h^0}(-k,t)\rangle +
\langle\widehat{y}(k,t) \widehat{y}(-k,t)\rangle.
\label{Sk_noise_fluct}
\end{equation}
In fact, the numerical data presented in Fig.~\ref{fig9} confirm that both components are
actually uncorrelated. Figure \ref{fig9} shows the structure factor of the faceted pattern
$h^0(x,t)$, the local fluctuations $y(x,t)$, and the complete interface $h(x,t)$ at two
different times. In particular, the data shown correspond to model B, but we obtained
identical results for for the other two models.

It becomes apparent that the anomalous scaling of the surface stems from that of the
faceted structure. Since the vertical scale in Fig.~\ref{fig9} is logarithmic, the sum in
Eq.~(\ref{Sk_noise_fluct}) will essentially be dominated by the largest of the two terms,
as one can easily see in Fig.~\ref{fig9}. Therefore, the large scale behavior of the
complete interface is totally dominated by the scaling behavior of the pattern component.
On the other hand, the structure factor $S(k,t)$ should crossover to that of the local
fluctuations, $\langle\widehat{y}(k,t) \widehat{y}(-k,t)\rangle$, at short wavelengths.
The structure factor of the pattern component shifts downwards with time because
$\alpha<\alpha_\mathrm{s}$. This is not the case for the structure factor of the local
fluctuations, which scale according to Family-Vicsek's, where $\alpha=\alpha_\mathrm{s}
\approx 0.45$ is fulfilled. Thus the length scale at which the local fluctuations dominate
the structure factor should grow with time. Whether the asymptotic scaling behavior of the
facets can be observed in the limit $t\rightarrow\infty$ for a finite system depends on
the relative strength of the local fluctuations. 

\section{Lyapunov vectors} So far we have studied the dynamics of random infinitesimal
perturbations, which is equivalent to study the dynamics of the most unstable direction in
tangent space, {\it i.~e.}~the first LV. We now devote this section to briefly describe
the dynamics of further unstable directions that growth at slower rates.

Let us start by introducing a few key quantities that shall be used in our analysis. The
evolution equation for an initially random infinitesimal perturbation,
Eq.~(\ref{evoerrors}), can be written in vectorial form as $\mathbf{\delta
u}(t+1)=\mathbf{J}[\mathbf{u}(t)]\mathbf{\delta u}(t)$, where $\mathbf{J}[\mathbf{u}(t)]$
is just the jacobian evaluated on the trajectory $\mathbf{u}(t)$ at time $t$. Since
infinitesimal perturbations evolve linearly there exists a linear operator
$\mathbf{M}(t_2,t_1)$ such that $\mathbf{\delta u}(t_2)=\mathbf{M}(t_2,t_1)\mathbf{\delta
u}(t_1)$, given by $\mathbf{M}(t_2,t_1) = \prod_{t_i=t_1}^{t_2-1}
\mathbf{J}[\mathbf{u}(t_i)]$. Oseledec's theorem \cite{Oseledec1968} implies that the
(symmetric) limit operator $\lim_{t_2 \rightarrow \infty} [\mathbf{M}(t_2,t_1)
\mathbf{M}^*(t_2,t_1)]^{1 /2(t_2-t_1)}$, where $\mathbf{M}^*$ denotes the adjoint of
$\mathbf{M}$, does exist and the logarithms of its eigenvalues are the Lyapunov exponents
$\lambda_n$ ($n =1, 2, \cdots, L$) and the eigenvectors are the so-called {\it backward}
Lyapunov vectors $\mathbf{b}_n$~\cite{eckmann,legras96}. In what follows, we will consider
standard ordering of the Lyapunov exponents,
$\lambda_1\geq\lambda_2\geq\ldots\geq\ldots\lambda_L$, and the corresponding vectors. We
use standard numerical techniques to calculate Lyapunov exponents and vectors
\cite{benettin_lyapunov_1980,GEIST1990}.

\begin{figure}
\centerline{\includegraphics *[width=0.40\textwidth]{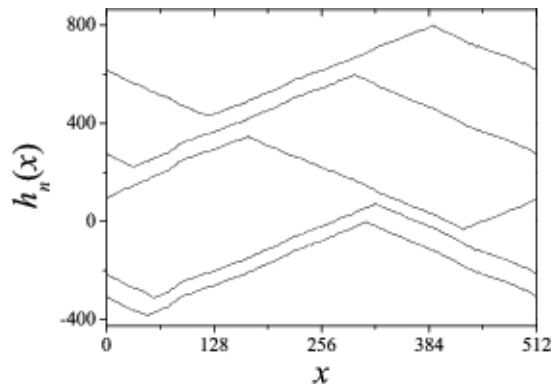}}
\caption{The surfaces $h_n(x)$ corresponding to the first five
Lyapunov-vectors are plotted (from bottom $n=1$ to top $n=5$).
Note that all the interfaces converge to a triangular structure.
The data presented were obtained for model B and have been shifted for better
visibility.\label{fig10}}
\end{figure}

As we have discussed in Secs.~III and IV, in inhomogeneous systems the main LV, the one
corresponding to the most unstable direction, converges to a faceted structure
asymptotically dominated by one single facet in the limit $t\rightarrow \infty$. In
Fig.~\ref{fig10} we plot the LV surfaces $h_n(x,t)=\ln(|b_n(x,t)|)$, $n\in\{1,2,3,4,5\}$,
corresponding to the first five backward LVs for model B at some arbitrary time $t$ during
evolution in a system of 512 coupled maps. The surface profiles in this plot are
arbitrarily shifted in the vertical direction to aid comparison. In the long time limit
the interface $h_1(x,t)$, the first LV surface, naturally coincides with the interface
$h(x,t)$ corresponding to an arbitrary initially random perturbation. It should be
observed that not only the first LV, but each surface $h_n(x,t)$ corresponding to the
$n$-th LV indeed converges to a structure dominated by a single triangle as well. This is
actually the case for all the inhomogeneous models studied in this paper.

Interestingly, the location of the cusp of the triangles, where the LVs spatially
localize, is not arbitrary. It is interesting to compare the positions of the global
maxima of the interface profiles with the values of the quenched disorder at those sites.
For model C, {\it i.~e.}~the RT case, the site where the first LV takes its maximum value
coincides with the site $x$ at which the coupling $\epsilon_x$ takes its lowest value,
the site where the second LV takes its maximum coincides with the site where
$\epsilon_x$ takes its second lowest value, and so forth. This is easy to interpret:
since errors get trapped at sites with lower diffusion, the result is that the LVs
strongly localize and utterly get frozen at those sites.

A very similar correlation effect between error localization and disorder can be seen for
model B, {\it i.~e.}~the RB case. Note that in the case of RB disorder the diffusion
coefficient at a given site $x$ is the sum $\epsilon_{x+1}+\epsilon_x$ of the local
diffusive couplings connecting $x$ to any of the two neighboring sites. In fact, we find
that the site where the first LV takes its maximum coincides with the site $x$ where the
diffusivity $\epsilon_{x+1}+\epsilon_x$ takes its lowest value, the site where the
second LV takes its maximum value coincides with the site where
$\epsilon_{x+1}+\epsilon_x$ takes its second lowest value and so forth. Thus, the
on-site growth velocity is again controlled by the probability for the perturbation to get
{\em trapped} at that site.

One would expect that a similar association of LVs maxima with disorder extrema should
also exist for model A. However, things are not that straightforward in this case. We
recall that in model A the disorder is introduced in the map constants and not in the
diffusive couplings, which are identical along the system. We think that due to the
dominance of small scale fluctuations up to quite large length scales for model A (see
Fig.~\ref{fig5}) as compared with the situation for models B and C, the maxima get
smoothed out and an extended neighborhood, rather than just one single site, controls the
final position where the LVs get frozen at long times. In other words, local
synchronization of neighboring sites over short scales makes it difficult to simply
identify the region where, at a coarse-grained level, the vectors will get asymptotically
trapped in the case of disorder of type A.

\section{Torward a phenomenological stochastic field theory} The evolution of
perturbations in homogeneous SECSs is known to be described in statistical terms by the
multiplicative stochastic equation
\begin{equation}
\frac{\partial}{\partial t}\delta u =  \partial_{xx}\delta u + \xi(x,t) \delta u,
\label{ruido}
\end{equation}
where $\xi(x,t)$ is a noise term that accounts for the chaotic fluctuations along
the trajectory and is assumed to be delta-correlated in space and time,
$\langle \xi(x,t) \, \xi(x',t') \rangle = 2 \, \sigma \, \delta(x-x') \, \delta(t-t')$.
Pikovsky and Kurths~\cite{pikovsky_roughening_1994} and Pikovsky and
Politi~\cite{pikovsky_dynamic_1998} proposed this equation as the proper candidate for
modeling the dynamics of the first Lyapunov vector from a statistical perspective. They
showed that it actually reproduces the main statistical properties of SECS in a variety of
systems~\cite{pikovsky_dynamic_1998}. This equation mimics the linear equations in tangent
space for the dynamics of infinitesimal perturbations in spatio-temporal chaotic systems.

An important feature of Eq.~(\ref{ruido}) is that, under a Hopf-Cole transformation,
$h=\ln|\delta u|$, it maps into the KPZ equation:
\begin{equation}
\frac{\partial}{\partial t} h = (\partial_x h)^2 + \partial_{xx} h + \xi(x,t),
\label{kpz}
\end{equation}
which ultimately justifies why the log-transformed (main) Lyapunov vector of many
spatiotemporal chaotic systems scales in space and time as a KPZ
surface~\cite{pikovsky_roughening_1994,pikovsky_dynamic_1998,pikovsky_dynamic_2001,
sanchez_rare_2004,szendro_CLV07}. Interestingly, Eq.~(\ref{ruido}) also appears in the
context of the classical problem of the directed polymer in a random
potential~\cite{kardar_zhang} (also see Ref.~\cite{halpin-zhang} for an detailed review).

We now explore extensions of this type of Langevin equations to the case of disordered
systems. A phenomenological stochastic field theory of this kind should be helpful in
order to identify the essential symmetries and mechanisms that determine the dynamics of
perturbations in chaotic disordered systems. The existence of scale invariance strongly
encourages this approach. Therefore, microscopic details of the models are expected to be
irrelevant at a sufficiently coarse-grained scale, as occurs in the case of homogeneous
systems.

In the case of Model A, Eq.\ (\ref{evoerrors}) can be considered as a straightforward
discrete version of
\begin{equation}
\frac{\partial}{\partial t} \delta u = \partial_{xx}[\partial_u f(x,u)\delta u]
+ [\partial_u f(x,u)-1] \, \delta u.\nonumber
\end{equation}
It is important to note that the noisy term $\partial_u f(x,u) \equiv \partial_y
f(x,y)\vert_{u(x,t)}$ depends implicitly on time through the trajectory $u(x,t)$ and so,
it includes both an annealed and a quenched component. The origin of the latter is the
inhomogeneous character of the coupled elements: Sites are more or less chaotic depending
on their position $x$ in the system. For instance, in the particular case of the logistic
map we have $\partial_u f(x,u) = c(x)[1-2c(x)u]$, where $c(x)$ is the local map parameter.
We argue that in the hydrodynamic limit the quenched part of the noise dominates over the
annealed one. The reason being that the quenched character of the disorder generates
long-range temporal correlations at each site, which are expected to control the dynamics
over the short-range correlations of the annealed random terms. 

These arguments suggest that for type A models the long time limit of main LV should be
generically described by a multiplicative Langevin equation like
\begin{equation}
\frac{\partial}{\partial t} \delta u =\partial_{xx} [(\zeta(x)\delta u] +
\partial_{xx}\delta u + \zeta(x)\delta u,\label{eq_model_A}
\end{equation}
where $\zeta(x)$ represents the quenched disorder. This equation can be further simplified
since the disordered diffusion term $\partial_{xx} [(\zeta(x)\delta u]$ is irrelevant as
compared with $\partial_{xx}\delta u$. To be precise, in the case of weak disorder that we
study here the inhomogeneous diffusion term simply renormalizes to an effective constant
diffusion term~\footnote{On the contrary, in the case of strong disorder the existence of
strongly trapping sites gives a sub-diffusive contribution to the dynamic
exponent~\cite{haus_diffusion_1987,bouchaud_anomalous_1990}. Again, this term would be
irrelevant in the long time limit, which is dominated by the constant diffusion term in
Eq.~(\ref{eq_model_A}).}. Finally we arrive at our stochastic field theory proposal for
model A type of systems:
\begin{equation}
\partial_t \delta u = D \partial_{xx} \delta u + \zeta(x)\delta u,\label{zhang}
\end{equation}
where $D$ is an effective diffusion constant and $\zeta(x)$ is a quenched columnar
disorder with delta correlations.

Interestingly, Eq.~(\ref{zhang}) appears in the context of diffusion in random
trapping/amplifying quenched potentials
\cite{zhang_diffusion_1986,ebeling_diffusion_1984,engel_comment_1987,tao_exact_1988,
tao_exact_1989,tao_path-integral_1991,rosenbluth_random_1989,valle_diffusion_1991}.
After the Hopf-Cole transform of the perturbation field, $\delta u = \exp(\mu h/D)$, one
obtains
\begin{equation}
\partial_t h(x,t) = \partial_{xx} h + \mu (\partial_x h)^2 + \zeta(x),\label{facets}
\end{equation}
which is a close relative of the KPZ equation, but here the additive noise term $\zeta(x)$
is quenched and delta correlated, $\langle\zeta(x) \zeta(x')\rangle = 2\theta \delta(x -
x')$. This equation has attracted some interest in the past, as it also describes the free
energy of directed and undirected polymers of length $t$ in the presence of columnar
disorder \cite{CATES1988,nattermann_diffusion_1989,krug1993,szendro_localization_2007}.

Only very recently~\cite{szendro_localization_2007}, it has been realized that the surface
described by Eq.~(\ref{facets}) exhibits anomalous scaling exponents induced by the
self-organization in a faceted structure. In Ref.~\cite{szendro_localization_2007} we
studied Eq.~(\ref{facets}) by means of extensive simulations and determined the critical
exponents to be $z=1.35\pm 0.05$, $\alpha=1.05\pm 0.05$ and $\alpha_\mathrm{s}=1.50\pm
0.05$. These values are in excellent agreement with the values reported here for the
critical exponents of the main LV surface in SECSs of type A (see Fig.~\ref{fig5}). This
strongly supports our claim that, in fact, Eq.~(\ref{zhang}) is a minimal model that
captures the essential ingredients dominating the evolution of errors in inhomogeneous
models of type A~\footnote{It is worth to mention here that there are analytical
results~\cite{krug1993} concerning Eq.~(\ref{facets}) showing that the correlation lenght
scales as $\sim t/(\ln t)^{3/4}$ for a Gaussian disorder, while slightly different
logarithmic corrections are obtained for different disorder distributions. This implies
that $z \approx 1.35$ has to be seen as an effective exponent, and as such will not be
universal. Nonetheless, these corrections are very weak and extremely difficult to
distinguish from purely power-law growth, even with several decades of scaling. We kindly
redirect the interested reader to Ref.~\cite{szendro_localization_2007} for further
details.}.

As it has already been discussed in Ref.~\cite{szendro_localization_2007}, the
introduction of an additional annealed noise term in Eq.~(\ref{facets}) is irrelevant in
the renormalization group sense, but leads to an increase of the typical size over which
the local random fluctuations (discussed in Sec.~V) dominate. Hence, for a finite size
system the asymptotic scaling behavior may not be observable if the thermal noise
amplitude is too large. This is of some importance for the situations we are interested in
because, when considering the evolution of perturbations in SECS, there will always be
some temporally fluctuating noise as explained above. Thus, even if quenched disorder is
present in the studied system, it may be very weak in comparison with the temporally
fluctuating one. In that case the behavior described here would only be observable in very
large systems.

Moreover, it should be noticed that, although the roughness exponents $\alpha$,
$\alpha_\mathrm{loc}$ and $\alpha_\mathrm{s}$ seem to be the same within error bars for
the three models of disorder studied in this paper, the dynamic exponent $z$ does take
different values. This suggests that the disorder symmetries that distinguish models A, B
and C are indeed relevant and lead to different nonequilibrium universality classes. One
can try to formulate a stochastic-field evolution equation similar to
Eq.~(\ref{eq_model_A}) for models B and C, but where the disorder enters in a different
form into the diffusion and the multiplicative term in order to reflect the different
symmetries of the coupling in these two models. Unfortunately, we have not been able to
find the correct equations reproducing the scaling behavior of the LV surfaces for
disordered systems in the class B and C. Similar arguments might be used
to write down the stochastic equation
\begin{equation}
\partial_t \delta u = \partial_{x}[\zeta(x)\partial_{x} \delta u] + \xi(x,t)\delta u,
\label{eq_model_B}
\end{equation}
for model B and
\begin{equation}
\partial_t \delta u = \partial_{xx}[\zeta(x) \delta u] + \xi(x,t)\delta u,
\label{eq_model_C}
\end{equation}
for model C, where $\xi(x,t)$ is a noise term. We have investigated the scaling properties
of Eqs.~(\ref{eq_model_B}) and (\ref{eq_model_C}) by means of numerical simulations and
found that, while these simple stochastic models generically lead to faceted surfaces with
the expected roughness exponents, they are unable to reproduce the correct dynamic
exponent of disordered SECS in the class B nor C. We believe that the reason behind this
disagreement lies in the existence of correlations between the quenched and annealed
fluctuating terms that are not incorporated in this naive approach. These non-trivial
correlations might also explain the, at least at first sight, counter intuitive result
that different dynamic exponents are observed in SECS of class B and C. Some remarks
regarding this point are now in order. 

Classical diffusion in random media involves the study of equations
like~(\ref{eq_model_B}) and (\ref{eq_model_C}), but where the multiplicative term
$\xi(x,t)\delta u$ is absent. It is well-known~\cite{bouchaud_anomalous_1990} that, in the
long time limit, dynamics of RB and RT models is identical, even for strong disorder.
Despite the different symmetries involved, the large scale physics is the same. However,
our numerical results indicate that model B (RB symmetry) and model C (RT symmetry) in
SECS seemingly belong to different universality classes. We claim that the multiplicative
noise, which is generically coupled to the quenched disorder in the systems we have
studied, breaks the duality between RB and RT type of disorder in this case, leading to
distinct dynamical exponents for the two different diffusion configurations.

\section{Conclusions} We have studied the evolution of infinitesimal perturbations in
inhomogeneous spatially extended systems exhibiting space-time chaos. Inhomogeneity is
introduced by means of a quenched disorder. We have considered one-dimensional coupled-map
lattices as a simple and computationally convenient model system to analyze some aspects
of chaos in the presence of disorder. In this regard, three different classes of models
have been investigated by means of extensive numerical simulations. In all cases we find a
strong localization behavior characterized by an exponential spatial profile around some
localization centers. In the long time limit the perturbation concentrates around one
single final attracting center. We have also studied the second, third, and so on backward
LVs corresponding to most rapidly expanding directions in tangent space and found a
correspondence between the localization centers and the positions in space corresponding
to increasing minima of the diffusion in the particular disorder realization.

Note that the strong localization behavior described here is essentially different from
the dynamic localization observed in the case of homogeneous SECS. In the latter,
perturbations do also localize on just a few sites, but the position of these sites keeps
fluctuating in time. However, in the presence of quenched disorder the sites
where perturbations localize are fixed by the corresponding realization of the disorder.

Moreover, by a standard mapping (Hopf-Cole transform) of the perturbation into a growing
surface we found that the LV associated surfaces self-organize in a faceted structure, at
variance with what occurs in homogeneous (non disordered) systems where one generically
obtains a surface in the universality class of KPZ. Interestingly, this faceted surface
was found to exhibit coarsening and anomalous kinetic roughening in agreement with
previous theoretical predictions~\cite{ramasco_generic_2000} for this type of
scale-invariant structures.

The evolution of infinitesimal perturbations in spatially extended chaotic systems with
quenched disorder may be described at a coarse-grained level as a diffusion process in a
random potential. Note, however, that the value of the dynamical exponent $z$ depends on
how disorder is introduced. The reason for this is possibly the influence of the disorder
distribution on the temporal dependence of the coarsening, as described in
Ref.~\cite{krug1993} in the context of diffusion (see
also Ref.~\cite{szendro_localization_2007} for a discussion relevant to the present work) 

We think that strong localization and anomalous scaling should play a role, {\it e.~g.}~in
realistic weather models, where quenched disorder is included in the form of inhomogeneous
boundary conditions representing certain geographical and topographical conditions.

\begin{acknowledgements} We thank D. Paz{\'o} for enlightening discussions on space-time
chaos and M. A. Mu{\~n}oz for his support and comments on this work. Financial support
from the Ministerio de Educaci\'on y Ciencia (Spain) under projects FIS2006-12253-C06-04
and CGL2007-64387/CLI is acknowledged.
\end{acknowledgements}


\begin{thebibliography}{10}

\bibitem{bohr}
T. Bohr, M.~H. Jensen, G. Paladin, and A. Vulpiani, {\em Dynamical Systems
  Approach to Turbulence} (Cambridge, Cambridge, 1988).

\bibitem{frisch}
U. Frish, {\em Turbulence} (Cambridge University Press, Cambridge, 1995).

\bibitem{kalnay}
E. Kalnay, {\em Atmospheric Modeling, Data Assimilation and Predictability}
  (Cambridge University Press, Cambridge, 2003).

\bibitem{eckmann}
J.-P. Eckmann and D. Ruelle, Rev. Mod. Phys. {\bf 57},  617  (1985).

\bibitem{ott}
E. Ott, {\em Chaos in Dynamical Systems} (Cambridge University Press,
  Cambridge, 1993).

\bibitem{legras96}
B. Legras and R. Vautard,  in {\em Predictability Vol. I}, ECWF Seminar, edited
  by T. Palmer (ECMWF, Reading, UK, 1996), pp.\ 135--146.

\bibitem{pikovsky_roughening_1994}
A.~S. Pikovsky and J. Kurths, Phys. Rev. E {\bf 49},  898  (1994).

\bibitem{pikovsky_dynamic_1998}
A. Pikovsky and A. Politi, Nonlinearity {\bf 11},  1049  (1998).

\bibitem{pikovsky_dynamic_2001}
A. Pikovsky and A. Politi, Phys. Rev. E {\bf 63},  036207  (2001).

\bibitem{sanchez_rare_2004}
A.~D. S\'anchez, J.~M. L\'{o}pez, M.~A. Rodr\'{i}guez, and M.~A. Mat\'{i}as,
  Phys. Rev. Lett. {\bf 92},  204101  (2004).

\bibitem{szendro_CLV07}
I.~G. Szendro, D. Paz\'o, M.~A. Rodr{\'\i}guez, and J.~M. L\'opez, Phys. Rev. E
  {\bf 76},  025202  (2007).

\bibitem{kardar1986}
M. Kardar, G. Parisi, and Y.-C. Zhang, Phys. Rev. Lett. {\bf 56},  889  (1986).

\bibitem{charney}
J.~G. Charney,  in {\em Dynamic Meteorology}, edited by P. Morel (Reidel,
  Boston, 1973), pp.\ 97--351.

\bibitem{pedlosky}
J. Pedlosky, {\em Geophysical Fluid Dynamics} (Springer-Verlag, Berlin, 1979).

\bibitem{legras_ghil}
B. Legras and M. Ghil, J. Atmos. Sci. {\bf 42},  433  (1985).

\bibitem{brainman}
Y. Brainman, J.~F. Linder, and W.~L. Ditto, Nature {\bf 378},  465  (1995).

\bibitem{wiesenfield}
K. Wiesenfield, P. Colet, and S. Strogatz, Phys. Rev. Lett. {\bf 76},  404
  (1996).

\bibitem{mousseau}
N. Mousseau, Phys. Rev. Lett. {\bf 77},  968  (1996).

\bibitem{ahlers_zillmer_pikovsky}
V. Ahlers, R. Zillmer, and A. Pikovsky, Phys. Rev. E {\bf 63},  036213  (2001).

\bibitem{haus_diffusion_1987}
J.~W. Haus and K.~W. Kehr, Phys. Rep. {\bf 150},  263  (1987).

\bibitem{bouchaud_anomalous_1990}
J.-P. Bouchaud and A. Georges, Phys. Rep. {\bf 195},  127  (1990).

\bibitem{krug1993}
J. Krug and T. Halpin-Healy, J. Phys. I France {\bf 3},  2179  (1993).

\bibitem{ramasco_generic_2000}
J.~J. Ramasco, J.~M. L\'{o}pez, and M.~A. Rodr\'{i}guez, Phys. Rev. Lett. {\bf
  84},  2199  (2000).

\bibitem{barabasi_fractal_1995}
A.-L. Barab{\'a}si and H.~E. Stanley, {\em Fractal Concepts in Surface Growth}
  (Cambridge University Press, Cambridge, 1995).

\bibitem{yang_instability_1994}
H.~N. Yang, G.~C. Wang, and T.~M. Lu, Phys. Rev. Lett. {\bf 73},  2348  (1994).

\bibitem{jeffries_instability_1996}
J.~H. Jeffries, J.-K. Zuo, and M.~M. Craig, Phys. Rev. Lett. {\bf 76},  4931
  (1996).

\bibitem{huo_anomalous_2001}
S. Huo and W. Schwarzacher, Phys. Rev. Lett. {\bf 86},  256  (2001).

\bibitem{santamaria_scaling_2002}
J. Santamaria {\it et~al.}, Phys. Rev. Lett. {\bf 89},  190601  (2002).

\bibitem{lopez_anomalous_1998}
J.~M. L\'{o}pez and J. Schmittbuhl, Phys. Rev. E {\bf 57},  6405  (1998).

\bibitem{morel_anomalous_1998}
S. Morel, J. Schmittbuhl, J.~M. L\'{o}pez, and G. Valentin, Phys. Rev. E {\bf
  58},  6999  (1998).

\bibitem{soriano_anomalous_2002}
J. Soriano {\it et~al.}, Phys. Rev. Lett. {\bf 89},  026102  (2002).

\bibitem{amar_groove_1993}
J.~G. Amar, P.-M. Lam, and F. Family, Phys. Rev. E {\bf 47},  3242  (1993).

\bibitem{schroeder_scaling_1993}
M. Schroeder {\it et~al.}, Europhys. Lett. {\bf 24},  563  (1993).

\bibitem{das_sarma_kinetic_1994}
S.~D. Sarma, S.~V. Ghaisas, and J.~M. Kim, Phys. Rev. E {\bf 49},  122  (1994).

\bibitem{das_sarma_scale_1996}
S.~D. Sarma, C.~J. Lanczycki, R. Kotlyar, and S.~V. Ghaisas, Phys. Rev. E {\bf
  53},  359  (1996).

\bibitem{smilauer_crossover_1994}
P. Smilauer and M. Kotrla, Phys. Rev. B {\bf 49},  5769  (1994).

\bibitem{bhattacharjee_infrared_1996}
J.~K. Bhattacharjee, S.~D. Sarma, and R. Kotlyar, Phys. Rev. E {\bf 53},  R1313
   (1996).

\bibitem{dasgupta_controlled_1996}
C. Dasgupta, S.~D. Sarma, and J.~M. Kim, Phys. Rev. E {\bf 54},  R4552  (1996).

\bibitem{lopez_power_1997}
J.~M. L\'{o}pez, M.~A. Rodr\'{i}guez, and R. Cuerno, Physica A {\bf 246},  329
  (1997).

\bibitem{lopez_superroughening_1997}
J.~M. L\'{o}pez, M.~A. Rodr\'{i}guez, and R. Cuerno, Phys. Rev. E {\bf 56},
  3993  (1997).

\bibitem{krug_turbulent_1994}
J. Krug, Phys. Rev. Lett. {\bf 72},  2907  (1994).

\bibitem{lopez_scaling_1999}
J.~M. L\'{o}pez, Phys. Rev. Lett. {\bf 83},  4594  (1999).

\bibitem{lopez_scaling_2005}
J.~M. L\'{o}pez, M. Castro, and R. Gallego, Phys. Rev. Lett. {\bf 94},  166103
  (2005).

\bibitem{Oseledec1968}
V. Oseledec, Trans. Moscow Math. Soc. {\bf 19},  179  (1968).

\bibitem{benettin_lyapunov_1980}
G. Benettin, L. Galgani, A. Giorgilli, and J.-M. Strelcyn, Meccanica {\bf 15},
  9  (1980).

\bibitem{GEIST1990}
K. Geist, U. Parlitz, and W. Lauterborn, Progress of theoretical physics {\bf
  83},  875  (1990).

\bibitem{kardar_zhang}
M. Kardar and Y.~C. Zhang, Phys. Rev. Lett. {\bf 58},  2987  (1987).

\bibitem{halpin-zhang}
T. Halpin-Healy and Y.~C. Zhang, Phys. Rep. {\bf 254},  215  (1995).

\bibitem{zhang_diffusion_1986}
Y.~C. Zhang, Phys. Rev. Lett. {\bf 56},  2113  (1986).

\bibitem{ebeling_diffusion_1984}
W. Ebeling, A. Engel, B. Esser, and R. Feistel, Journal of Statistical Physics
  {\bf 37},  369  (1984).

\bibitem{engel_comment_1987}
A. Engel and W. Ebeling, Phys. Rev. Lett. {\bf 59},  1979  (1987).

\bibitem{tao_exact_1988}
R. Tao, Phys. Rev. Lett. {\bf 61},  2405  (1988).

\bibitem{tao_exact_1989}
R. Tao, Phys. Rev. Lett. {\bf 63},  2695  (1989).

\bibitem{tao_path-integral_1991}
R. Tao, Phys. Rev. A {\bf 43},  5284  (1991).

\bibitem{rosenbluth_random_1989}
M.~N. Rosenbluth, Phys. Rev. Lett. {\bf 63},  467  (1989).

\bibitem{valle_diffusion_1991}
A. Valle, M.~A. Rodr\'{i}guez, and L. Pesquera, Phys. Rev. A {\bf 43},  2070
  (1991).

\bibitem{CATES1988}
M.~E. Cates and R.~C. Ball, J. Phys. {\bf 49},  2009  (1988).

\bibitem{nattermann_diffusion_1989}
T. Nattermann and W. Renz, Phys. Rev. A {\bf 40},  4675  (1989).

\bibitem{szendro_localization_2007}
I.~G. Szendro, J.~M. L\'{o}pez, and M.~A. Rodr\'{i}guez, Phys. Rev. E {\bf 76},
   011603  (2007).

\end{thebibliography}
\end{document}